# A REPRESENTATION OF CHANGES OF IMAGES AND ITS APPLICATION FOR DEVELOPMENTAL BIOLOLOGY

Gene Kim and MyungHo Kim

Bioinformatics Frontier Inc.

93 B Taylor Ave

East Brunswick, NJ 08816

mkim@biofront.biz

**Abstract**. In this paper, we consider a series of events observed at spaced time intervals and present a method of representation of the series. To explain an idea, by dealing with a set of gene expression data, which could be obtained from developmental biological experiments, the procedures are sketched with comments in some details. We mean representation by choosing a proper function, which fits well with observed data of a series, and turning its characteristics into numbers, which extract the intrinsic properties of fluctuating data. With help of a machine learning techniques, this method will give a classification of developmental biological data as well as any varying data during a certain period and the classification can be applied for diagnosis of a disease.

## §1. Introduction

This work was motivated by questions of several medical/biological scientists who I had chances to work with. Here are some of projects have been discussed:

1. Changes of gene expression rates in the brains of mouse with/without electric shock for equally spaced time intervals
2. Association study between changes of gene expression rates and radiation sensitivity
3. Changes of clinical tests including international prostate symptom score (IPSS) while a series of dose of some medicine are applied

They are typical everlasting problems in the biological/medical world. These days, to attack them, people are using the techniques such as DNAchip, gene expression, micro-array analysis developed recently. However, despite of those techniques being so useful and powerful, the statistical method for interpreting and getting something meaningful from the data is not that successful as expected. This might be due to lack of understanding the problems.

Though those three problems look different, the intrinsic properties are the same conceptually. Here we introduce a universal method, which may put those three problems into a single setting, which is ready for use of machine-learning methods such as SVMs.

Our approach could lead to a new paradigm of not only genetic research

(including developmental research), but also diagnosis such as disease progress. Throughout this paper, for convenience, we will consider the case of a single gene expression data, though it could be easily applied to other problems as well as multiple genes.

DNAchip and micro-array techniques turn the expression rates into densities of stained images, which may be recorded as a series of numbers. All the experiments and phenomenon tell us that the numbers fluctuate over the time and if we plot them as a graph, it is similar to a portion of the following shape:

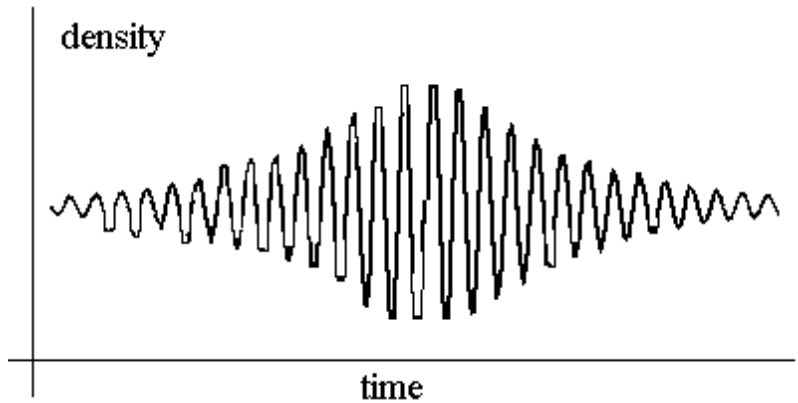

Fig. 1

So to speak, the expression rates will start to increase slowly, reach a peak, and decrease to a certain level. As the next steps for describing this phenomenon, it is required to find a proper function that will fit with it.

§2. Representation

In this section, we will describe the procedures of representing a set of gene expression data into a vector can be described. Each step will be given explanation in some details.

## Step 1. Fixing a candidate function

Since we need a periodic-looking, fluctuating functions, it will be good to start with *sin*(*t*) or cos(*t*), while, for increasing and decreasing effects, the exponential function would be the feasible choice. Consequently, a possible function for representing the changes of the expression rate would be of the form *exp*(*kt*)*sin*(*mt*) or *exp*(*kt*)*cos*(*mt*), where *k* and *m* are real constants. The exponential functions appear everywhere in science, especially, in modeling problems and theories and it is no wonder we see it here once again. We all see some functions keep being brought up and exponential function is one of them. Though it is so familiar that we feel so comfortable, once in a while, one might ask of oneself the following question: why is this so important, appear so often and works well? There must be a simple explanation for this, but I did not see the reason why so many people use some specific functions or at least I could not find any people explain to me. Nevertheless, I would like to justify my choice of the exponential function here. The first clue is in the profound experimental fact, i.e., that the radioactive decay is measured in terms of *half-life* – the number of years required for half of the atoms in a sample of radioactive material to decay. Mathematically this is expressed as

$$y' = ky$$

Here *y* represents the mass and *k* is a constant. Then the general type of a solution looks like *y* = *Cexp*(*kt*), where *t* is a time and *C* is a constant. Second, every object is made of atoms, bulbs, engines, living cells, enzymes, DNA, RNA so on, and, as we deal in this paper, in most of cases, it might be used to observe some sort of *life expectancy* of a certain phenomenon or behavior.

Here are a few examples of functions possible for fitting, depending on the nature and behavior of observed data.

Examples

1. $f(t) = C_1 \exp(k_1 t)\sin(m_1 t)$ for $-a \leq t \leq 0$

    $C_2 \exp(k_2 t)\sin(m_2 t)$ for $0 \leq t \leq a$

2. $f(t) = C_1 \exp(k_1 t)\sin(m_1 t + \varphi_1)$ for $-a \leq t \leq 0$

    $C_2 \exp(k_2 t)\sin(m_2 t + \varphi_2)$ for $0 \leq t \leq a$

    where $C_2 \sin\varphi_1 = C_2 \sin\varphi_2$

3. $f(t) = -m_1 t + b_1,\ 0 \leq t \leq \dfrac{b_1}{m_1}$

    $m_2 t + b_2,\ \dfrac{b_2}{m_2} \leq t \leq \dfrac{b_3 - b_2}{m_2 + m_3}$

    $-m_3 t + b_3,\ \dfrac{b_3 - b_2}{m_2 + m_3} \leq t \leq \dfrac{b_3}{m_3}$

    $\vdots \qquad \vdots$

    $m_{2k} t + b_{2k},\ \dfrac{b_{2k}}{m_{2k}} \leq t \leq \dfrac{b_{2k+1} - b_{2k}}{m_{2k} + m_{2k+1}}$

    $-m_{2k+1} t + b_{2k+1},\ \dfrac{b_{2k+1} - b_{2k}}{m_{2k} + m_{2k+1}} \leq t \leq \dfrac{b_{2k+1}}{m_{2k+1}}$

    ,where $\dfrac{b_{2i-1}}{m_{2i-1}} = \dfrac{b_{2i}}{m_{2i}}$, for each $i = 1,..,k$.

where *a, k's, m's* are positive real numbers.

### Step 2. Determining coefficients, *C's* and *k's*

Once we fix a candidate function, it remains to determine the coefficients *C's* and *k's* etc., for each set of data. This may be achieved by using the least square sum principle with high accuracy set to our own standard. Commercial software such SAS and SPSS etc. are available for such calculation, namely, R-squared. The least square sum method is, as the most popular one for fitting a curve/function with experimental data, to find coefficients of a function of given type, by minimizing the sum of square of errors. More precisely, given a set of data points, $(x_1, y_1)$, $(x_2, y_2)$... $(x_n, y_n)$ and a candidate function $f$ with undetermined coefficients, the unknowns in $f$ would be determined so that the summation of errors, given by

$$\sum_{i=1}^{n}(f(x_i) - y_i)^2,$$

be minimized. As in the case of choosing a candidate, why do we have to use two? Why not $\sum_{i=1}^{n}|f(x_i) - y_i|$, $\sum_{i=1}^{n}(f(x_i) - y_i)^3$ or $\sum_{i=1}^{n}(f(x_i) - y_i)^4$ for measuring the degree of errors? For odd number like three, the sum cancels each other and does not reflect our purpose, i.e., the degree of errors. When the exponent is two, it is the smallest and good for further manipulation, i.e., we could use many tools, calculus, involving differentiation unlike the absolute value function, | |.

### Step 3. Representing each set of data as a vector

From the first two steps, we obtained a functional representation for observed data, i.e., a function fitting with the data. Consequently, with respect to the fixed type, each function is represented as a set of coefficients calculated in the step two. Suppose we observed a single gene expression rate of an object, *A*, for a certain fixed period and, for simplicity, we assume for a moment that the data fits well with model,

$$y = C\exp(kt)\sin(mt)$$

, where *y* is the expression rate. Then we can say that the set of numbers (*C, k, m*) represent the object, *A*. In other words, the object may be identified with the triple (*C, k, m*), so are people with their own social security numbers. For the general case, i.e., the gene expression rates of *n* multiple genes, we will get $(C_1, k_1, m_1)$, $(C_2, k_2, m_2)$,..., $(C_n, k_n, m_n)$, which form a vector in the 3*n* dimensional Euclidean space.

Once again, we are in a familiar position, ready to apply the support vector machine (or a machine learning tools such as neural network and decision tree etc.) to find a criterion for separating one from another. (For more details, see [1], [2], [3] and [4]).

§3. Discussions

The method in the previous section can be applied for observed events during a certain period. Reactions of drugs, any changes of substance in changing environments etc. could be classified and used for a criterion of diagnosis of a disease or analysis of statuses.

On the matter of choosing a function of a proper type, besides the exponential function, in quantitative analysis of science including genetics, often we encounter functions of linear type, *ax+b* where *x* variable, *a* and *b* constants. In many cases, the linear functions or, sometimes, quadratic functions are simply assumed without any explanations. For readers who are in a hurry to grasp the rest of the story, it is hardly expected to spend time in getting a convincing answer for that. I believe that we could find an answer to this on one hypothesis and a famous mathematical theorem of calculus.

1. Hypothesis

   Any quantitative measurement can be expressed as a function of some variables.

2. Taylor theorem

   Roughly speaking for a function of a variable, any smooth function *f* may be expressed as follows;

   $f(x) = f(0) + f'(0)x + f^{(2)}(0)/2! x^2 + f^{(3)}(0)/3! x^3 + \ldots\ldots + f^{(n)}(0)/n! x^n + R_n(x)$

   where $R_n(x) = f^{(n+1)}(z)/(n+1)! x^{n+1}$

The variables in the hypothesis are not observable with modern technology or maybe impossible for human, but theoretically, we can assume there are such variables. For the last several decades, we witnessed the rapid development of computers, super computing powers. This luxury enable us to raise the power of exponents, in other words, we do not need to restrict ourselves to function of degree one or two. This might be a bit wild, but what if we allow singularities in the quantitative function, then we might have to replace Taylor by the Maclaurin

theorem.(See for [5])

## Acknowledgement

We are grateful to Profs Chul Ahn of University of Texas at Houston for encouraging comments and criticism. We also give special thanks to Young-Joon Hong, M.D. at Korean Cancer Center Hospital and Prof. Chulwoo Kim at medical school of Seoul National University for invaluable discussions.

Interface of Experimental and Computational Modeling, 2002 (http://dimacs.rutgers.edu/Workshops/Complexity/program.html), submitted for patent (US Application No: 10/336,334)